\documentclass[journal]{IEEEtran}
\usepackage{amsmath,amsfonts}
\usepackage{algorithmic}
\usepackage{array}
\usepackage[caption=false,font=normalsize,labelfont=sf,textfont=sf]{subfig}
\usepackage[group-separator={,},group-minimum-digits=3]{siunitx}
\usepackage{textcomp}
\usepackage{stfloats}
\usepackage{url}
\usepackage{verbatim}
\usepackage{graphicx}
\usepackage{balance}
\usepackage{mathrsfs}
\begin{document}
\title{ \fontsize{20pt}{\baselineskip}\selectfont
Illegal Intelligent Reflecting Surface Based Active Channel Aging: When Jammer Can Attack Without Power and CSI}
\author{
{Huan~Huang,~\textit{Student~Member,~IEEE,}
Ying~Zhang,
Hongliang~Zhang,~\textit{Member,~IEEE,}
Chongfu~Zhang,~\textit{Senior~Member,~IEEE,}
and~Zhu~Han,~\textit{Fellow,~IEEE}}
\thanks{This work was supported by the National Key R\&D Program of China (2018YFB1801302). (\textit{Corresponding author: Chongfu Zhang})}
\thanks{H. Huang, Y. Zhang, and C. Zhang are with the School of Information and Communication Engineering, University of Electronic Science and Technology of China, Chengdu 611731, China (e-mail: hhuang@std.uestc.edu.cn; yzhang1@std.uestc.edu.cn; cfzhang@uestc.edu.cn).}
\thanks{H. Zhang is with the School of Electronics, Peking University, Beijing 100871, China (email: hongliang.zhang92@gmail.com).}
\thanks{Z. Han is with the University of Houston, Houston, TX 77004, USA (email: zhan2@uh.edu).}
}

\maketitle

\begin{abstract}
Illegal intelligent reflecting surfaces (I-IRSs), i.e., the illegal deployment and utilization of IRSs, impose serious harmful impacts on wireless networks. The existing I-IRS-based illegal jammer (IJ) requires channel state information (CSI) or extra power or both, and therefore, the I-IRS-based IJ seems to be difficult to implement in practical wireless networks.
To raise concerns about significant potential threats posed by I-IRSs, we propose an alternative method to jam legitimate users (LUs) without relying on the CSI.
By using an I-IRS to actively change wireless channels, the orthogonality of multi-user beamforming vectors and the co-user channels is destroyed, and significant inter-user interference is then caused, which is referred to as active channel aging.
Such a fully-passive jammer (FPJ) can launch jamming attacks on multi-user multiple-input single-output (MU-MISO) systems via inter-user interference caused by active channel aging, where the IJ requires no additional transmit power and instantaneous CSI. 
The simulation results show the effectiveness of the proposed FPJ scheme. Moreover, we also investigate how the transmit power and the number of quantization phase shift bits influence the jamming performance.
\end{abstract}


\section{Introduction}
\IEEEPARstart{D}{ue} to the intrinsic characteristics of wireless channels, i.e., broadcast and superposition, wireless networks are vulnerable to jamming attacks (also referred as to interference attacks), and it is difficult to protect transmitted signals from unauthorized recipients~\cite{PLSsur1}. Intelligent reflecting surfaces (IRSs) has been an emerging wireless technology for 5G, 6G and beyond~\cite{IRSsur1,IRSsur2}. Legitimate IRSs can be used to provide an important approach for enhancing the physical layer security (PLS) in wireless networks~\cite{ResGroup,IRSPLSadd}.

Therefore, many previous studies have investigated the use of legitimate IRSs to improve PLS~\cite{ANIRS,ActiveJammerIRS}. In~\cite{ANIRS}, IRSs combined with artificial noise (AN) or friendly jamming at the access point (AP) are used for security enhancement in the presence of illegal eavesdroppers. In~\cite{ActiveJammerIRS}, the authors proposed an IRS-assisted anti-jamming scheme against jamming attacks, where a friendly IRS is used to prevent the illegal jammer (IJ) from jamming legitimate users (LUs). Note that the legitimate AP in the legitimate IRS aided scenario knows the legitimate IRS's information, like its location, and can control the reflecting phase shifts of the legitimate IRS.

In contrast, illegal IRSs (I-IRSs) represent the illegal deployment and utilization of IRSs~\cite{IIRSSur}, where the legitimate AP does not know the I-IRSs' information
and also can not control the I-IRSs.
Due to the passive nature, the I-IRSs are hard to be detect. Consequently, the I-IRSs impose a more serious harmful impact on PLS. For example, an I-IRS has been employed to deteriorate signals at LUs in the presence of jamming attacks~\cite{IIRSSur}, where the I-IRS aggravates the AN generated by the IJ to reduce the received signal-to-noise ratio (SNR) or the signal-to-interference-noise ratio (SINR). However, there are two requirements in existing methods to achieve the I-IRS-based IJ.

\textit{1) I-IRSs need to know the channel state information (CSI) of all channels involved.} 
Yet, the uplink channel estimation for IRS-aided channels remains difficult due to the passive nature of IRSs~\cite{DaipartI}.
Acquiring the I-IRS-aided channels' CSI at IJ is too idealistic to implement in practice.
Although illegal jamming can be achieved without the CSI by broadcasting the AN~\cite{ANwoCSI}, the performance gain obtained by implementing an I-IRS, in this case, is limited as reflecting phase shifts of the I-IRS are hard to optimize without the CSI.

\textit{2) A large amount of power is needed to transmit jamming signals continuously.} 
Even a few papers attempt to realize an I-IRS-based passive jammer (PJ) without the transmit power 
for single-user systems~\cite{PassJamSU}, which minimizes the received power at the LU by destructively adding the signal from the AP-IRS-User channel.
However, this I-IRS-based PJ still requires the CSI of IRS-aided channels to optimize the I-IRS's reflecting phase shifts.

Limited by these two requirements above, especially the CSI acquisition, the I-IRS-based IJ seems to be difficult to implement in practical wireless networks. So in this paper, we try to answer the following research question: \textit{{Can IJs jam LUs without both the transmit power and the CSI?}}

To draw attention to the impact of I-IRSs on multi-user multiple-input single-output (MU-MISO) systems, we propose an I-IRS-based fully-passive jammer (FPJ) that can launch jamming attacks without relying on the transmit power and the CSI. To the best of our knowledge, it is the first time that an IJ can jam LUs without the CSI. 
\begin{itemize}
\item An I-IRS is exploited to actively change wireless channels, and therefore, the orthogonality of the multi-user active beamforming vectors and the co-user channels is destroyed, which is referred to as \textbf{active channel aging}$\footnote{Channel aging is CSI inaccuracy due to time variation of wireless channels and delays in the computation~\cite{ChanAge}. In this work, we actively introduce CSI inaccuracy by using an I-IRS. To differentiate, we call it active channel aging.}$.
\item During \textit{the reverse pilot transmission (RPT) phase}, we randomly generate reflecting phase shifts for the I-IRS. During \textit{the data transmission (DT) phase}, we randomly generate other reflecting phase shifts. The I-IRS acts like a \textbf{``disco ball"} without optimizing its phase shifts based on the CSI. The resulting serious inter-user interference due to active channel aging jams the LUs effectively.
\end{itemize}

\emph{Notation:} We use bold capital type for a matrix, e.g., $\boldsymbol{\Phi}$, small bold type for a vector, e.g., $\boldsymbol{\varphi}$, and italic type for a scalar, e.g., $K$. Moreover, the superscripts $(\cdot)^{H}$ and $(\cdot)^{T}$ denote the Hermitian transpose and the transpose. Moreover, the symbols $|\cdot|$ and $\|\cdot\|$ denote the absolute value and the Frobenius norm.
\section{System Statement}
In this section, first, we describe the general mode of an MU-MISO system jammed by the I-IRS-based FPJ. Then, we give the optimization metric and state the two communications phases: \emph{the RPT phase} and \emph{the DT phase}.
\subsection{System Model and Channel Model}
\begin{figure}[!t]
\centering
\includegraphics[scale=0.65]{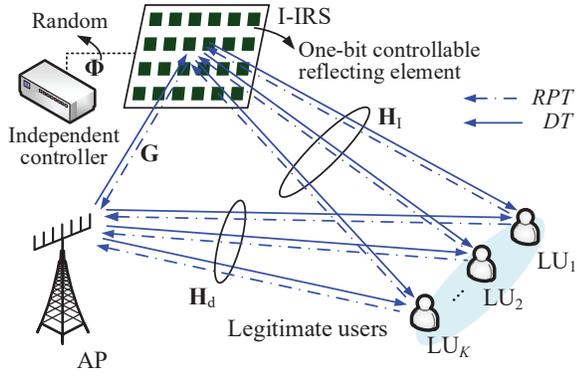}
\caption{Illustration of a MU-MISO system jammed by the I-IRS-based FPJ, where phase shifts of the I-IRS are randomly generated by the independent I-IRS controller. RPT: reverse pilot transmission; DT: data transmission.}
\label{fig1}
\end{figure}

Figure~\ref{fig1} schematically illustrates a MU-MISO system jammed by the I-IRS-based FPJ, where the legitimate AP is equipped with an $N_{\rm A}$-element uniform linear array (ULA) and communicates with $K$ single-antenna LUs termed ${\rm {LU}}_{1}, {\rm {LU}}_{2}, \cdots, {\rm {LU}}_{K}$. An I-IRS comprised of $N_{\rm I}$ one-bit controllable reflecting elements is deployed near the AP$\footnote{Based on the existing literature on the IRS's deployment location~\cite{IRSdeployment}, the IRS should be deployed as close to users or as close to the AP as possible to increase its effect. Yet, in the jamming scenario, we make the more robust assumption that the IJ does not know any information about LUs, for instance, LUs' locations and CSI. Therefore, we deploy the I-IRS near the AP.}$ to jam LUs.
When the data signal $s_k\in \mathbb{C}$ for ${\rm{LU}}_k$ $(1\ \leq  k \leq K)$ is normalized to unit power, the signal received at ${\rm{LU}}_k$ is expressed as,
\begin{equation}
{y_k} = \boldsymbol{h}_{{\rm{com}},k}^H\!\sum_{u = 1}^K {{\boldsymbol{w}_u}{s_u}}  + {n_k},
\label{eq1}
\end{equation}
where $\boldsymbol{h}^H_{\!{\rm{com}},k}\! = \!\left( \!{ {\boldsymbol{h}_{{\rm{I}},k}^H{{\bf{\Phi }}}{{\bf{G}}}} \!+\! \boldsymbol{h}_{{\rm{d}},k}^H} \right) \!\in\! {\mathbb{C}^{1\!\times\!{N_{\rm{A}}}}}$ denotes the combined channel between the legitimate AP and ${\rm{LU}}_k$, $\boldsymbol{h}_{{\rm{I}},k} \in\mathbb{C}^{N_{\rm{I}}\!\times\! 1}$ denotes the channel between the I-IRS and ${\rm{LU}}_k$, ${\bf{G}} \in\mathbb{C}^{N_{\rm{I}}\!\times\! N_{\rm{A}}}$ denotes the channel between the legitimate AP and the I-IRS, and $\boldsymbol{h}_{{\rm{d}},k} \in\mathbb{C}^{N_{\rm{A}}\!\times\! 1}$ denotes the direct channel between the legitimate AP and ${\rm{LU}}_k$.

In \eqref{eq1}, $\boldsymbol{\Phi}\!=\!{\rm{diag}}(\boldsymbol{\varphi})\in\mathbb{C}^{N_{\rm{I}}\!\times\! N_{\rm{I}}}$ represents the reflecting matrix of the I-IRS, where the one-bit reflecting vector ${\boldsymbol{\varphi}}$ is expressed as ${\boldsymbol{\varphi}} = \left[{e^{j{\varphi_{1}}}}, \cdots, {e^{j{\varphi_{{N\!_I}}}}}\right]^H$, and ${\varphi_{n}}\in \Omega  = \left\{0, \pi \right\} (1\le n \le N_{\rm I})$ denotes reflecting phase shift of the $n$-th reflecting element. The independent I-IRS controller
generates ${\boldsymbol{\varphi}}$ and then controls the I-IRS to implement the corresponding phase shifts.
Besides, $\boldsymbol{w}_k$ 
denotes the active beamforming at the AP for ${\rm{LU}}_k$, and $n_k$ denotes the additive white Gaussian noise with 0 mean and $\sigma^2$ variance, i.e., $n_k\sim \mathcal{CN}\left(0,\sigma^2\right)$.

For ease of representation, we further define the multi-user direct channel between the AP and the LUs, the multi-user channel between the I-IRS and the LUs, as well as the multi-user combined channel between the AP and all LUs as ${\bf{H}}_{\rm{d}}^H \!=\!\left[{\boldsymbol{h}}_{{\rm{d}},1},{\boldsymbol{h}}_{{\rm{d}},2},\cdots,{\boldsymbol{h}}_{{\rm{d}},K}\right]^H$, ${\bf{H}}_{{\rm{I}}}^H \!=\!\left[\boldsymbol{h}_{{\rm{I}},1},{\boldsymbol{h}}_{{\rm{I}},2},\cdots,\boldsymbol{h}_{{\rm{I}},K}\right]^H$, and ${\bf{H}}_{{\rm{com}}}^H=\left[ {{\boldsymbol{h}_{{\rm{com}},1}},{\boldsymbol{h}_{{\rm{com}},2}},\cdots,{\boldsymbol{h}_{{\rm{com}},K}}} \right]^H$, respectively. Furthermore, the multi-user active beamforming at the AP is denoted as ${\bf{W}}\!=\!\left[{\boldsymbol{w}}_1,{\boldsymbol{w}}_2,\ldots,{\boldsymbol{w}}_K\right]$.

The multi-user direct channel ${\bf{H}}_{\rm{d}}$ follows Rayleigh fading, while the IRS-aided channels ${\bf{G}}$ and $\boldsymbol{h}_{{\rm{I}},k}$ follow Rician fading~\cite{AORIS}. Specifically, ${\bf{G}}$ and $\boldsymbol{h}_{{\rm{I}},k}$ are modeled as
\begin{alignat}{1}
&{{\bf{G}}}\! = \!{\mathscr{L}}_{\rm G} \!\left(\!\!{\sqrt {\frac{{{\kappa_{\rm \!G}}}}{{1\!+\!{\kappa_{\rm \!G}}}}} {\bf{G}}^{{\rm{LOS}}}\!+\!\sqrt {\frac{1}{{1 + {\kappa_{\rm \!G}}}}} {\bf{G}}^{{\rm{NLOS}}}}\!\!\right),\notag\\
&{\boldsymbol{h}_{{\rm{I}},k}}\! = \!{\mathscr{L}}_{{\rm I},k} \!\left(\!\!{\sqrt {\frac{{{\kappa_{{\rm{I}}}}}}{{1\!+\!{\kappa _{{\rm{I}}}}}}} \boldsymbol{h}_{{\rm{I}},k}^{{\rm{LOS}}}\!+\!\sqrt {\frac{1}{{1\!+\!{\kappa_{{\rm{I}}}}}}}\!\boldsymbol{h}_{{\rm{I}},k}^{{\rm{NLOS}}}}\!\!\right),
\label{eq2}
\end{alignat}
where ${\mathscr{L}}_{\rm G}$ and ${\mathscr{L}}_{{\rm I},k}$ represent the large-scale path loss between the AP and the I-IRS and that between the I-IRS and ${\rm{LU}}_k$, and ${\kappa _{{\rm{G}}}}$ and ${\kappa _{{\rm{I}}}}$ are the Rician factors of ${{\bf{G}}}$ and ${\boldsymbol{h}_{{\rm{I}},k}}$.

In \eqref{eq2}, ${\bf{G}}^{{\rm{LOS}}}$ and $\boldsymbol{h}_{{\rm{I}},k}^{{\rm{LOS}}}$ are the line-of-sight (LOS) components of ${{\bf{G}}}$ and ${\boldsymbol{h}_{{\rm{I}},k}}$, and ${\bf{G}}^{{\rm{NLOS}}}$ and $\boldsymbol{h}_{{\rm{I}},k}^{{\rm{NLOS}}}$ are non-line-of-sight (NLOS) components. The NLOS components follow Rayleigh fading, while the LOS components are~\cite{AORIS},
\begin{alignat}{1}
&{\bf{G}}^{{\rm{LOS}}} = \sqrt {{N_{\rm{I}}}{N_{\rm A}}} {\boldsymbol{\alpha} _{\rm{I}}}\left( {{\vartheta},{\theta}} \right)\boldsymbol{\alpha} _{\rm{A}}^H\left( {{\phi}} \right),\notag\\
&\boldsymbol{h}_{{\rm{I}},k}^{{\rm{LoS}}} = \sqrt {{N_{\rm{I}}}} {\boldsymbol{\alpha} _{\rm{I}}}\left( {{\vartheta _{{\rm{I}},k}},{\theta_{{\rm{I}},k}}} \right),
\label{eq3}
\end{alignat}
where $\boldsymbol{\alpha} _{\rm{A}}$ and $\boldsymbol{\alpha} _{\rm{I}}$ are the array responses~\cite{AORIS}.
\subsection{Wireless Communications: The RPT and DT Phases}\label{CommPhase}
In practice, the main aim of a MU-MISO system is to maximize a certain performance metric that generally is a strictly-increasing utility function of SINR~\cite{BF}. Specifically, a widely-used performance metric is the sum rate, which is expressed as ${R_{{\rm{sum}}}} = \sum\nolimits_{k = 1}^K {{R_k}}  = \sum\nolimits_{k = 1}^K {{{\log }_2}\left( {1 + {\gamma _k}} \right)}$. According to~\eqref{eq1}, the received SINR $\gamma_k$ at ${\rm{LU}}_k$ is stated as,
\begin{equation}
{\gamma _k} = \frac{{{{\left| {\boldsymbol{h}_{{\rm{com}},k}^H{\boldsymbol{w}_k}} \right|}^2}}}{{\sum\limits_{u \ne k} {{{\left| {\boldsymbol{h}_{{\rm{com}},k}^H{\boldsymbol{w}_u}} \right|}^2} + {\sigma ^2}} }}.
\label{eq6}
\end{equation}
\subsubsection{Acquiring CSI During The RPT Phase}
From \eqref{eq6}, it can be seen that the optimization of multi-user active beamforming  ${\bf{W}} = \left[{\boldsymbol{w}}_1,{\boldsymbol{w}}_2,\ldots,{\boldsymbol{w}}_K\right]$ at the AP aims to maximize the signal term $\left| {\boldsymbol{h}_{{\rm{com}},k}^H{\boldsymbol{w}_k}} \right|$ while minimizing the inter-user interference term ${\sum\nolimits_{u \ne k}{\left| {\boldsymbol{h}_{{\rm{com}},k}^H{\boldsymbol{w}_u}} \right|}}$. In order to optimize ${\bf{W}}$, the CSI of $\bf{H}_{{\rm{com}}}$ must be obtained at the AP$\footnote{In the MU-MISO system under I-IRS-based jamming attacks, it is impractical to acquire the CSI of IRS-aided channels and the direct channel, respectively. The legitimate AP cannot know any information about the I-IRS, like its location, much less jointly train the IRS-based channels with the I-IRS. Namely, the legitimate AP can only obtain the CSI of $\bf{H}_{{\rm{com}}}$. Note that the CSI of $\bf{H}_{{\rm{com}}}$ is easily obtained at the legitimate AP when ${\bf{\Phi}}$ is determined, which is the traditional MISO channel estimation. The phase shifts of the I-IRS are generated at random by the independent I-IRS controller, and therefore, ${\bf{\Phi}}$ is always determined for the legitimate AP, as shown in Fig.~\ref{fig1}.}$.
Generally, the CSI can be acquired during \textit{the RPT phase} according to the pilot estimation, as shown in Fig.~\ref{fig1}. More specifically, to acquire the CSI of $\boldsymbol{h}_{{\rm{com}},k}$, the ${\rm{LU}}_k$ sends pilot signals to the legitimate AP, and the AP then estimates $\boldsymbol{h}_{{\rm{com}},k}$ by certain traditional solutions, for instance, the least square (LS) algorithm~\cite{DaipartI}.
\subsubsection{Precoding During The DT Phase}
Based on the obtained CSI in \textit{the RPT phase}, the multi-user active beamforming used during \textit{the DT phase} can be designed. Generally, the multi-user active beamforming optimization problem is a nondeterministic polynomial-time (NP)-hard problem, and therefore, computing the optimal multi-user active beamforming is difficult. To this end, some heuristic beamforming designs, which can achieve near-optimal performance, have been investigated.

A widely known beamforming solution is the zero-forcing beamforming (ZFBF) algorithm~\cite{BF}, which causes zero inter-user interference. Specifically, the multi-user active beamforming ${\bf{W}}_{\rm{ZF}}$ computed via the ZFBF algorithm is written as
\begin{equation}
{{\bf{W}}_{{\rm{ZF}}}} = \frac{{{{\bf{H}}_{{\rm{com}}}}{{\left( {{\bf{H}}_{{\rm{com}}}^H{{\bf{H}}_{{\rm{com}}}}} \right)}^{ - 1}}{{\bf{P}}^{ \frac{1}{2}}}}}{{{{\left\| {{{\bf{H}}_{{\rm{com}}}}{{\left( {{\bf{H}}_{{\rm{com}}}^H{{\bf{H}}_{{\rm{com}}}}} \right)}^{ - 1}}} \right\|}^2}}},
\label{eq7}
\end{equation}
where ${{\bf{P}}^{\frac{1}{2}}} = {\rm{diag}}\left( {\sqrt {{p_1}} ,\sqrt {{p_2}} , \cdots ,\sqrt {{p_K}} } \right)$, and $p_k$ represents the transmit power allocated to ${\rm{LU}}_k$. The power allocation must satisfy the constraint that $\sum\nolimits_{k = 1}^K {{p_k}}  \le {P_0}$, where $P_0$ is the total transmit power at the AP. The optimal power allocation can be calculated by the water-filling algorithm~\cite{BF}.
\subsubsection{Orthogonal Interference Subspace}
According to \eqref{eq6}, the ratio of inter-user interference to noise (I/N) ${\mathscr{I}}$ is equal to
\begin{equation}
{\mathscr{I}} = \sum\limits_{k = 1}^K \sum\limits_{u \ne k} \frac{{{{\left| {\boldsymbol{h}_{{\rm{com}},k}^H{\boldsymbol{w}_u}} \right|}^2} }}{\sigma^2} .
\label{eq8}
\end{equation}

Incorporating \eqref{eq7} into \eqref{eq8}, it is clear that ${\mathscr{I}} = 0$ due to the presence of the pseudoinverse ${{\left( {{\bf{H}}_{{\rm{com}}}^H{{\bf{H}}_{{\rm{com}}}}} \right)}^{ - 1}}$. In other words, ZFBF causes zero inter-user interference by projecting the user channel $\boldsymbol{h}_{{\rm{com}},k}$ onto the subspace that is orthogonal to the co-user channels $\boldsymbol{h}_{{\rm{com}},1},\cdots,\boldsymbol{h}_{{\rm{com}},k-1},\boldsymbol{h}_{{\rm{com}},k+1},\cdots,\boldsymbol{h}_{{\rm{com}},K}$, i.e., the orthogonal interference subspace.
\section{I-IRS-based Fully-Passive Jammer via Active Channel Aging}\label{PassJamm}
To raise concerns about the potential threat that an I-IRS could launch jamming attacks without the transmit power or even the CSI, we introduce a CSI-based PJ without the transmit power in Section~\ref{PJwCSI}, i.e., the extension of~\cite{PassJamSU}. Furthermore, the results from the CSI-based PJ are used as benchmarks.
In Section~\ref{proposed}, we propose an I-IRS-based FPJ via active channel aging. By destroying the orthogonality of the multi-user active beamforming vectors and the co-user channels, the proposed I-IRS-based FPJ can jam LUs without the transmit power and the CSI.
\subsection{CSI-Based Jamming Attacks Without Power}\label{PJwCSI}
To implement the extension of~\cite{PassJamSU}, it is necessary to consider the most ideal case for jamming attacks: the legitimate AP only knows the CSI of ${\bf{H}}_{\rm d}$ and then calculates the multi-user active beamforming ${\bf{W}}_{\rm{d}}$ via the ZFBF algorithm, while the independent I-IRS controller knows the CSI of ${\bf{H}}_{\rm{d}}$, ${\bf{H}}_{{\rm{I}}}$, and ${\bf{G}}$ as well as ${\bf{W}}_{\rm{d}}$.
The CSI-based PJ can launch jamming attacks without the transmit power, where the reflecting vector for the I-IRS is optimized by minimizing a certain performance metric. Taking the example of minimizing the sum rate $R_{\rm{sum}}$ received at LUs, the optimization of the one-bit reflecting vector is mathematically represented as
\begin{alignat}{1}
&\mathop {\min }\limits_{\boldsymbol{\varphi}}  {R_{{\rm{sum}}}} = \mathop {\min }\limits_{\boldsymbol{\varphi}}  \sum\limits_{k = 1}^K {{{\log }_2}}\!\! \left( {1 + \frac{{{{\left| {{\boldsymbol{h}}_{{\rm{com}},k}^H{{\boldsymbol{w}}_{{\rm{d}},k}}} \right|}^2}}}{{\sum\limits_{u \ne k} {{{\left| {{\boldsymbol{h}}_{{\rm{com}},k}^H{{\boldsymbol{w}}_{{\rm{d}},u}}} \right|}^2}}  + {\sigma ^2}}}} \right)\label{addeq81}\\
&{\rm{s}}.{\rm{t}}.\;\;{\varphi _n} \in \Omega, n = 1,2, \cdots ,{N_{\rm{I}}}.
\label{addeq82}
\end{alignat}

The phase shift optimization problem in~\eqref{addeq81} can be solved by enumerating all possible $\left\{ {{\varphi _n}} \right\}_{n = 1}^{{N_{\rm{I}}}}$ combinations.
However, there are $2^{N_{\rm I}}$ different combinations, and thus the computational complexity is large.

To this end, we first relax the discrete phase shift constraint in~\eqref{addeq82} to a continuous constraint. Mathematically, the reflecting vector optimization is relaxed to
\begin{alignat}{1}
& \mathop {\max }\limits_{\bar{\boldsymbol{\varphi}}} \! \sum\limits_{k = 1}^K \!-{{{\log }_2}}\!\! \left( \!{1 +\! \frac{{{{\left|\!{
\left({{{\bar{\boldsymbol{\varphi}}}{{\rm{diag}}(\boldsymbol{h}_{{\rm{I}},k}^H)}{{\bf{G}}}} \!+\! \boldsymbol{h}_{{\rm{d}},k}^H}\!\right)\!{{\boldsymbol{w}}_{{\rm{d}},k}}}\! \right|}^2}}}{{\sum\limits_{u \ne k} {{{\left| {\left(\!{{{\bar{\boldsymbol{\varphi}}}{{\rm{diag}}(\boldsymbol{h}_{{\rm{I}},k}^H)}{{\bf{G}}}} \!+ \! \boldsymbol{h}_{{\rm{d}},k}^H}\!\right)\!{{\boldsymbol{w}}_{{\rm{d}},u}}} \right|}^2}} \!\! +\! {\sigma ^2}}}}\! \right)\label{addeq911}\\
&{\rm{s}}.{\rm{t}}.\;\;{\bar {\varphi}_n} \in \left[ {0,2\pi } \right], n = 1,2, \cdots ,{N_{\rm{I}}}.
\label{addeq922}
\end{alignat}

The objective function in~\eqref{addeq911} is then a continuous and differentiable function of $ \bar {{\boldsymbol{\varphi}}}$, 
and the constraint in~\eqref{addeq922} creates a complex circle manifold. Therefore, the optimization problem in~\eqref{addeq911} can be computed by the Riemannian conjugate gradient (RCG) algorithm~\cite{AORCG}. After computing the continuous reflecting vector $\bar{\boldsymbol{\varphi}}$, the discrete reflecting vector is obtained by
\begin{alignat}{1}
&\mathop {\min }\limits_{{\boldsymbol{\varphi}}}  {\left\| {{{\boldsymbol{\varphi}}} - {\bar{\boldsymbol{\varphi}}} } \right\|^2} \label{addeq101}\\
\nonumber
&{\rm{s}}.{\rm{t}}.\;\;\eqref{addeq82}.
\end{alignat}

The complexity of the benchmarking CSI-based PJ is ${\mathcal O}\!\left(I_{\rm R}K^2N^2_{\rm I} \right)$, where $I_{\rm R}$ represents the iteration times of the RCG algorithm. In each iteration, the complexity comes mainly from calculating the Euclidean gradient~\cite{AORCG}. Specifically, the complexity of the Euclidean gradient calculation is ${\mathcal O}\!\left( K^2N^2_{\rm I} \right)$. Moreover, the complexity of the discreteization of $\bar{\boldsymbol{\varphi}}$ expressed by~\eqref{addeq101} is ${\mathcal O}\!\left( 2 N_{\rm I} \right)$. When the number of reflecting elements packed on the I-IRS is large ($N_{\rm I} \gg 1$), the complexity of the discreteization, i.e., ${\mathcal O}\!\left( 2 N_{\rm I} \right)$, can be ignored.
\subsection{I-IRS-Based Jamming Attacks Without Power and CSI}\label{proposed}
Although the CSI-based PJ proposed in Section~\ref{PJwCSI} can jam without the transmit power, the CSI of all channels needs to be obtained at the independent I-IRS controller, which is difficult to satisfy in practice. In wireless communications, the AP needs to obtain the CSI during \textit{the RPT phase} before \textit{the DT phase}, as stated in Section~\ref{CommPhase}.

\begin{figure}[!t]
\centering
\includegraphics[scale=0.6]{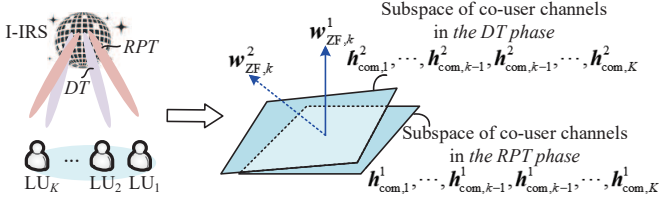}
\caption{I-IRS-based FPJ via {\textbf{active channel aging}}, where the I-IRS acts like a {\textbf{``disco ball"}} and ZFBF cannot project the user channel to the orthogonal interference subspace.}
\label{fig2}
\end{figure}

\subsubsection{The RPT Phase}
During \textit{the RPT phase}, the one-bit reflecting vector for the I-IRS is generated by tuning the $n$-th reflecting element to a random phase shift belonging to $\Omega$, i.e., ${{\varphi}}_n^1 \sim \mathcal{U}\left(\Omega\right)$. More particularly, the reflecting vector ${\boldsymbol{\varphi}}^1$ follows the uniform distribution denoted ${\boldsymbol{\varphi}}^1 \sim \mathcal{U}\left(\Omega^{N_{\rm I}}\right)$. It is worth noting that the independent I-IRS controller in the proposed scheme does not need to optimize the reflecting phase shifts of the I-IRS.

Consequently, the multi-user combined channel estimated by the AP is written as $({{\bf{H}}_{{\rm{com}}}^1} )^H = {\bf{H}}_{\rm I}^H{\rm{diag}}\left({\boldsymbol{\varphi}}^1\right){\bf{G}}+{\bf{H}}_{\rm d}^H = \left[{{\boldsymbol{h}}_{{\rm{com}},1}^1}, {{\boldsymbol{h}}_{{\rm{com}},2}^1}, \cdots, {{\boldsymbol{h}}_{{\rm{com}},K}^1}\right]^H$.
Based on ${{\bf{H}}_{{\rm{com}}}^1}$, the AP can compute the multi-user active beamforming used in \textit{the DT phase} that is expressed as
\begin{equation}
{\!{\bf{W}}_{{\!\rm{ZF}}}^1} \!=\! \frac{{\!{{\bf{H}}_{\!{\rm{com}}}^1}{{\left( {(\!{{\bf{H}}_{\!{\rm{com}}}^1} \!)^{\!H}{\!{\bf{H}}_{\!{\rm{com}}}^1}} \right)}^{\! - 1}}{\!{\bf{P}}^{\! \frac{1}{2}}}}}{{{{\left\|\! {{\bf{H}}_{\!{\rm{com}}}^1}{{\left( {(\!{{\bf{H}}_{\!{\rm{com}}}^1} \!)^{\!H}{\!{\bf{H}}_{\!{\rm{com}}}^1}} \right)}^{\! - 1}} \!\right\|}^2}}} \! = \!\left[{\!{\boldsymbol{w}}_{\!{\rm{ZF}},1}^1}, {\!{\boldsymbol{w}}_{\!{\rm{ZF}},2}^1}, \!\cdots\!, {\!{\boldsymbol{w}}_{\!{\rm{ZF}},\!K}^1} \!\right],
\label{ZF1}
\end{equation}
where ${{\boldsymbol{w}}_{{\rm{ZF}},k}^1}$ is orthogonal to  the subspace of co-user channels $\boldsymbol{h}_{{\rm{com}},1}^1,\cdots,\boldsymbol{h}_{{\rm{com}},k-1}^1,\boldsymbol{h}_{{\rm{com}},k+1}^1,\cdots,\boldsymbol{h}_{{\rm{com}},K}^1$.
\subsubsection{The DT Phase}
Then, during \textit{the DT phase}, the one-bit reflecting vector of the I-IRS is formed according to another reflecting vector ${\boldsymbol{\varphi}}^2$ that also follows the uniform distribution in $\Omega$, i.e., ${\boldsymbol{\varphi}}^2 \sim \mathcal{U}\left(\Omega^{N_{\rm I}}\right)$. Therefore, during \textit{the DT phase}, the multi-user combined channel is changed to
\begin{equation}
(\!{{\bf{H}}_{{\rm{com}}}^2} \!)^{\! H} \!\!=\! {\!\bf{H}}_{\rm I}^H{\!\rm{diag}}\!\left(\!{\boldsymbol{\varphi}}^2\!\right){\!\bf{G}}\!+{\!\bf{H}}_{\!\rm d}^H \!=\! \left[{\!{\boldsymbol{h}}_{\!{\rm{com}},1}^2}, {\!{\boldsymbol{h}}_{\!{\rm{com}},2}^2}, \!\cdots\!, {\!{\boldsymbol{h}}_{\!{\rm{com}},K}^2}\!\right]^{\! H}.
\label{addHcom2}
\end{equation}

Including~\eqref{ZF1} and~\eqref{addHcom2} into~\eqref{eq6}, the actual received SINR ${{\bar{\gamma}}_k}$ at ${\rm{LU}}_k$ during \textit{the DT phase} is
\begin{equation}
{{\bar{\gamma}}_k} = \frac{{{{\left| {({{\boldsymbol{h}}_{{\rm{com}},k}^2})^H{\boldsymbol{w}_{{\rm {ZF}},k}^1}} \right|}^2}}}{{\sum\limits_{u \ne k} {{{\left| {({{\boldsymbol{h}}_{{\rm{com}},k}^2})^H{\boldsymbol{w}_{{\rm {ZF}},u}^1}} \right|}^2} + {\sigma ^2}} }}.
\label{eq9}
\end{equation}

The complexity of our proposed scheme comes from randomly generating the two reflecting vectors used in \textit{the RPT phase} and \textit{the DT phase}, which is only  ${\mathcal O}\!\left( 2 N_{\rm I} \right)$. Compared with the benchmarking CSI-based PJ, the I-IRS's controller in the proposed I-IRS-based FPJ not only does not require the CSI of all channels involved, but also the complexity of the proposed I-IRS-based FPJ is much lower.
\subsubsection{Active Channel Aging}
Based on~\eqref{ZF1} and~\eqref{addHcom2}, the reflecting vectors for the I-IRS are different and random during \textit{the RPT phase} and \textit{the DT phase} (like a ``disco ball" shown in Fig.~\ref{fig2}), which destroys the orthogonality generated from ZFBF due to active channel aging.
The ${{\boldsymbol{w}}_{{\rm{ZF}},k}^1}$ is only orthogonal to the subspace of co-user channels $\boldsymbol{h}_{{\rm{com}},1}^1,\cdots,\boldsymbol{h}_{{\rm{com}},k-1}^1,\boldsymbol{h}_{{\rm{com}},k+1}^1,\cdots,\boldsymbol{h}_{{\rm{com}},K}^1$, and thus ${\mathscr{I}}$ in \eqref{eq8} is then equal to $\sum\nolimits_{k = 1}^K{\sum\nolimits_{u \ne k} \frac{{{{\left| {(\boldsymbol{h}_{{\rm{com}},k}^2)^H{\boldsymbol{w}^1_{{\rm{ZF}},u}}} \right|}^2} }}{\sigma^2}}$, which is no longer zero due to active channel aging. 

As a result, the actual received SINR ${{\bar{\gamma}}_k}$ in~\eqref{eq9} achieved under the proposed I-IRS-based FPJ is dramatically reduced compared to that without attacks.
We stated that the reflecting vector for the I-IRS is different during \textit{the RPT phase} and \textit{the DT phase}. In fact, there is no need for precise synchronization in practical implementation. Assuming that the periods of \textit{the RPT phase} and \textit{the DT phase} are $T_{\rm{r}}$ and $T_{\rm{d}}$ ($T_{\rm{r}} \le T_{\rm{d}}$), the reflecting vector changes randomly with a period of no more than $T_{\rm{r}}$, and active channel aging then occurs.
\section{Simulation Results and Discussion}\label{ResDis}
Consider a MU-MISO system with four single-antenna LUs, where the legitimate AP is equipped with a 12-element ULA~\cite{BF} and an I-IRS contains 1$,$024 reflecting elements ($N_{\rm{I},y}=N_{\rm{I},z}=32$).
Moreover, the AP is located at (0m, 0m, 0m) and the four LUs are randomly distributed in a circle centered at (200m, 0m, 0m) with a radius of 10m, while the I-IRS is deployed at (5m, 5m, 2m).

Most of the existing performance-enhancing IRS-aided systems make the assumption that ${\bf{H}}_{\rm d}$ has significant large-scale path loss or is blocked, while the large-scale path losses of $\bf{G}$ and ${\bf{H}}_{\rm I}$ are much smaller~\cite{AORIS,AORCG}. However, this assumption is too idealistic for jamming attacks.
According to the 3GPP propagation environment~\cite{3GPP}, the large-scale path losses ${\mathscr{L}}_{k}$, ${\mathscr{L}}_{\rm G}$ and ${\mathscr{L}}_{\rm{I},k}$ are set as ${\mathscr{L}}_{k}\!=\!32.6\!+\!22{\log _{10}}({d_{k}})$, ${\mathscr{L}}_{\rm G} \!=\!35.6\!+\!20{\log _{10}}({d_{{\rm{G}}}})$ and ${\mathscr{L}}_{{\rm I},k}\!=\!35.6\!+\!22{\log _{10}}({d_{{\rm I},k}})$, where $d_{k}$ is the distance between the AP and ${\rm{LU}}_k$, $d_{\rm G}$ is the distance between the AP and and the I-IRS, and ${d_{{\rm{I}},k}}$ is the distance between the I-IRS and ${\rm{LU}}_k$ $(1\!\leq\! k\!\leq\!4)$. Moreover, $\sigma^2\!=\!-170\!+\!10\log _{10}(BW)$ dBm, where $BW$ denotes the transmission bandwidth and $BW\!=\!180$ kHz~\cite{AORCG}.
We compare the proposed I-IRS-based FPJ with three benchmarks.

\textit{1) Benchmark 1:} The average sum rates without IJ (w/o IJ) are computed based on the multi-user direct channel, where the received SINR $\gamma_k$ at ${\rm{LU}}_k$ is $\gamma_k = \frac{{{{\left| {{\boldsymbol{h}}_{{\rm{d}},k}^H{\boldsymbol{w}_{{\rm{d}},k}}} \right|}^2}}}{{\sum\nolimits_{u \ne k} {{{\left| {\boldsymbol{h}_{{\rm{d}},k}^H{\boldsymbol{w}_{{\rm{d}},u}}} \right|}^2}} + {\sigma ^2}}}$.

\textit{2) Benchmark 2:} The average sum rates under the active jammer (w/ AJ/N) with different ratios of the jamming power to the noise power (AJ/N) at each LU. More specifically, the received SINR $\gamma_k$ at ${\rm{LU}}_k$ under active jamming is expressed as $\gamma_k = \frac{{{{\left| {{\boldsymbol{h}}_{{\rm{d}},k}^H{\boldsymbol{w}_{{\rm{d}},k}}} \right|}^2}}}{{\sum\nolimits_{u \ne k} {{{\left| {\boldsymbol{h}_{{\rm{d}},k}^H{\boldsymbol{w}_{{\rm{d}},u}}} \right|}^2}} + {P_J} + {\sigma ^2}}}$, where AJ/N = ${P_J/ {\sigma ^2}}$=5 dB and 10 dB, respectively.

\textit{2) Benchmark 3:} The CSI-based PJ in Section~\ref{PJwCSI}, i.e., the extension of~\cite{PassJamSU}.

\begin{figure}[!t]
\centering
\includegraphics[scale=0.68]{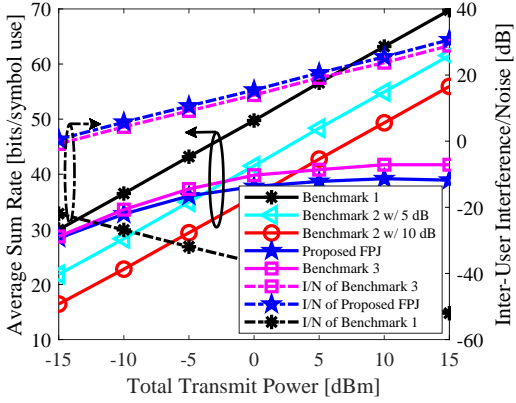}
\caption{Average sum rates (left, solid lines) and I/N (right, dash-dot lines) of different schemes vs total transmit power.}
\label{fig3}
\end{figure}

Fig.~\ref{fig3} illustrates the average sum rates via the proposed FPJ and the above three benchmarks, where ${\mathscr{I}}$ generated from them is also given. By destroying the orthogonality of the multi-user active beamforming vectors and the co-user channels, the inter-user interference becomes significant due to active channel aging. The reflecting vector in the proposed FPJ affects both the multi-user combined channel and the multi-user active beamforming, while the reflecting vector in the CSI-based PJ just impacts the multi-user combined channel. As shown in Fig.~\ref{fig3}, ${\mathscr{I}}$ from the proposed FPJ is more serious than that from the CSI-aided PJ (Benchmark 3). Therefore, the proposed FPJ can jam LUs without the transmit power and the CSI, even more effectively than the CSI-aided PJ.

From Fig.~\ref{fig3}, one can see that the sum rate of Proposed FPJ is smaller than that of Benchmark 2 with 5 dB AJ/N when the total transmit power is greater than 0 dBm. In contrast to the active jamming, the jamming launched by the proposed FPJ cannot be mitigated by increasing the total transmit power.

\begin{figure}[!t]
\centering
\includegraphics[scale=0.68]{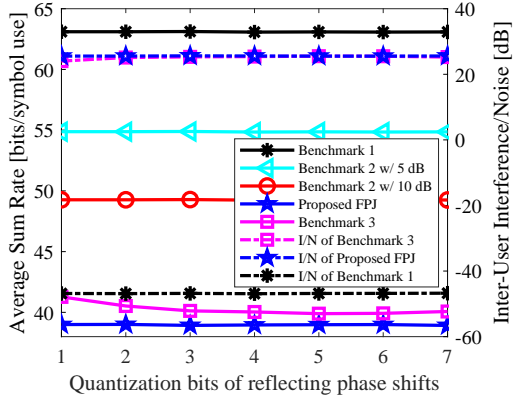}
\caption{Influence of quantization reflecting phase shift bits.}
\label{fig4}
\end{figure}

To show the influence of the number of quantization reflecting phase shift bits, the relationships between the average sum rates and quantization bits are illustrated in Fig.~\ref{fig4}. One can see that the proposed FPJ is robust to the quantization bits since the reflecting vector is randomly generated. Based on the proposed FPJ, the one-bit I-IRS is enough to launch effective jamming attacks on LUs.
The greater the number of quantization bits, the smaller the difference ${\left\| {{{\boldsymbol{\varphi}}} - {\bar{\boldsymbol{\varphi}}} } \right\|^2}$ in~\eqref{addeq101} is.
Although the sum rate achieved by Benchmark 3 decreases with the number of quantization bits, the high-bit I-IRS requires high physical implementation costs.

\begin{figure}[!t]
\centering
\includegraphics[scale=0.68]{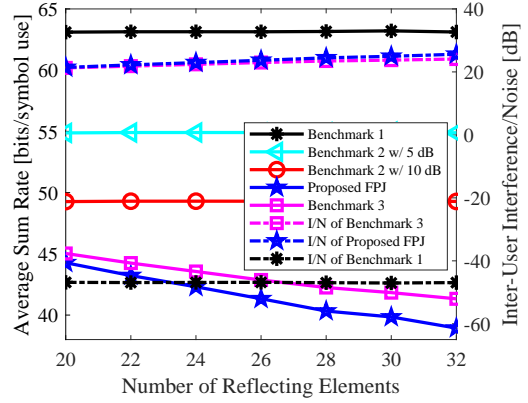}
\caption{Influence of the number of reflecting elements.}
\label{fig5}
\end{figure}

Moreover, Fig.~\ref{fig5} shows the relationship between the sum rates and the number of reflecting elements as well as that between I/N and the number of reflecting elements. The difference between the sum rates achieved by Benchmark 3 and Proposed FPJ increases with the number of reflecting elements. On the one hand, active channel aging becomes more significant with the number of reflecting elements, and thus the corresponding jamming attacks are more effective. On the other hand, the minimum value of ${\left\| {{{\boldsymbol{\varphi}}} - {\bar{\boldsymbol{\varphi}}} } \right\|^2}$ in~\eqref{addeq101}  gets bigger with the number of reflecting elements. In practice, an IRS generally consists of massive reflecting elements, which is beneficial to the proposed I-IRS-based FPJ.
\section{Conclusions}
In this letter, we investigated the impact of I-IRSs on MU-MISO systems, where an I-IRS-based FPJ was proposed. By exploiting an I-IRS to cause active channel aging, we have demonstrated that the proposed FPJ can jam without relying on the transmit power and the CSI.
Due to the impacts on both the multi-user combined channel and the multi-user active beamforming, the jamming launched by the proposed FPJ is even more effective than that launched by the CSI-aided PJ.
Meanwhile, the proposed FPJ is robust to the number of quantization reflecting phase shift bits. Different from the active jamming attacks, the jamming attacks launched by the proposed FPJ cannot be mitigated by increasing the total transmit power at the legitimate AP. When the legitimate AP has large transmit power, the proposed FPJ can jam LUs more effectively.
Moreover, the proposed FPJ can be perfectly hidden in wireless environments because it does not require additional transmit power and instantaneous CSI. 

\end{document}